\documentclass[12pt]{iopart}
\usepackage{amsmath}
\usepackage{amsfonts}
\usepackage{dcolumn}                             
\usepackage{amssymb}
\usepackage{graphicx,graphics,wrapfig,rotating}         
\usepackage[german,english]{babel}
\usepackage{bm,pstricks}                         

\newcommand{\rem}[1]{}

\newcommand{\ket}[1]{|\,#1\,\rangle}                %
\newcommand{\bra}[1]{\langle\,#1\,}                 %
\newfont{\Bb}{msbm10}                   %
\def\eea{\end{eqnarray}}
\def\bea{\begin{eqnarray}}
\def\ee{\end{equation}}
\def\be{\begin{equation}}
\def\nn{\nonumber}

\begin{document}

\title{Weyl law for fat fractals}

\author{Mar\'\i a E. Spina$^1$, Ignacio Garc\'\i a-Mata$^{1,2}$ and Marcos Saraceno$^{1,3}$}
\address{$^1$ Departamento de F\'\i sica, CNEA, Libertador 8250,
C1429BNP Buenos Aires, Argentina}
\address{$^2$ CONICET, Avda. Rivadavia 1917, CP C1033AAJ Buenos Aires, Argentina}
\address{$^3$ Escuela de Ciencia y Tecnolog\'ia, UNSAM, Alem 3901, B1653HIM Villa Ballester, Argentina1}
\ead{spina@tandar.cnea.gov.ar}
\date{\today}

\begin{abstract}
It has been conjectured that for a class of piecewise linear maps
the closure of the set of images of the discontinuity has the
structure of a fat fractal, that is, a fractal with positive
measure. An example of such maps is the sawtooth map in
the elliptic regime. 
In this work we analyze this problem quantum mechanically in the semiclassical regime. 
We find
that the fraction of states localized
on the unstable set satisfies a modified fractal Weyl law, where
the exponent is given by the exterior dimension of the fat
fractal.
\end{abstract}




\maketitle
\section{Introduction}
The general question of how, and at which rate, quantum states
explore phase spaces with a complex structure has drawn much
attention recently. In the case of chaotic repellers the ordinary
Weyl law of one quantum state per phase space cell $ (2 \pi
\hbar)^d $ is { substituted} by a conjectured fractal Weyl law  with the
integer dimension $d$ replaced by the fractal dimension of the
repeller \cite{chao}. In this paper we address this question with
respect to elliptic { quantum} maps with a discontinuity. These maps exhibit
a complex phase space where regular open islands coexist with the
set consisting of the iterations of the discontinuity. This
irregular set has a very complex structure and the question is
whether it supports a finite fraction of stationary states in the
semiclassical limit. This has been answered rigorously by the
negative in \cite{mark} with the assumption that the $d$-Minkowski
content of this set vanishes,  { i.e.} that { its} closure 
has zero Lebesgue measure. 

A paradigmatic example of such maps is the sawtooth map 
in the elliptic regime
\cite{las,ash}. In \cite{ash} Ashwin shows numerical evidence
indicating that for this map the irregular sea, i.e. the closure
of the images of the discontinuity, has fractal structure and
positive Lebesgue measure. An analogous result is presented in
\cite{sco} for the case of a discontinuous Hamiltonian mapping on
the sphere \cite{mil}, which is conjectured to produce circle
packing in phase space. Also in this case, the irregular sea
appears to have non zero measure. These subsets of fractal
structure and positive measure are loosely called `fat fractals'
and can be characterized by their exterior dimension and their
Lebesgue measure \cite{far}. 

{ In this work we  conjecture, and validated by numerical
evidence, that in the case where the irregular set is a `fat' fractal,
for which the condition in \cite{mark} is not met, 
a non vanishing fraction of
eigenstates is} {localized in} { it and the rate at which this is
achieved follows a generalized Weyl law governed by its exterior dimension.
In order to obtain the results presented, an effective method 
to count the eigenstates 
living in the irregular set is needed. We introduce a procedure which consists in opening
the map along the discontinuity line. The eigenstates supported by  the irregular set have 
a large overlap with the open area and thus the corresponding eigenvalues migrate to the interior of the unit disk.
By computing the fraction of these
short-living states as a function of the resolution, we are able to find
a scaling law, depending on the Lebesgue measure and exterior
dimension of the set.}

{ We organized our contribution as follows.
Our calculations will be performed for the the elliptic
sawtooth map.  We will first recall some properties of the
classical version of this map in section \ref{sec:classmap}. Then in 
section \ref{sec:qmap} we consider its quantized
version and using the poposed method we compute the fraction of eigenstates supported by the
irregular subset . Finally we give some concluding remarks in section \ref{sec:conc}}
\section{The classical sawtooth map}
\label{sec:classmap}
\begin{figure}[h]
\begin{center}
\includegraphics[width=0.9\linewidth]{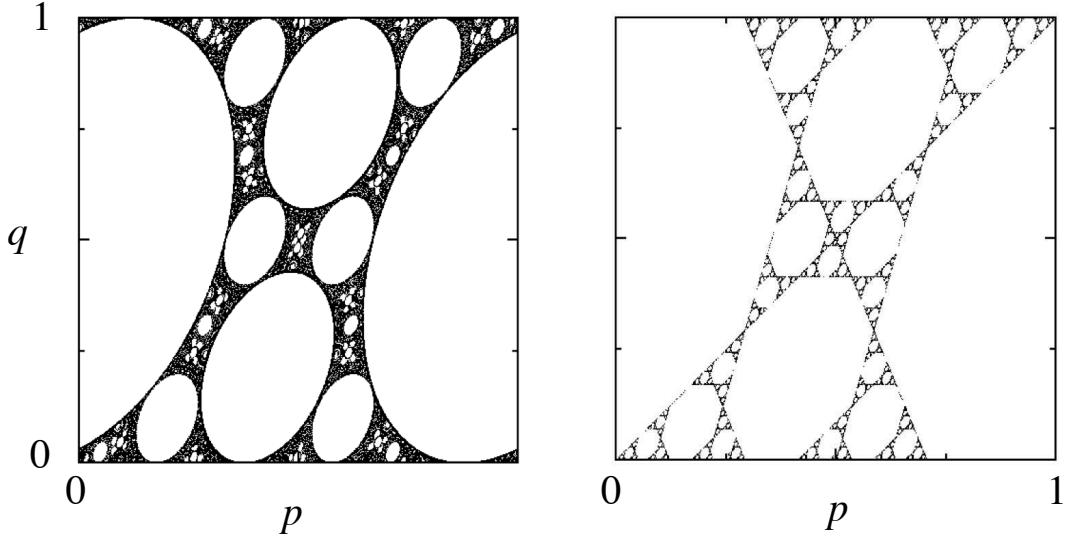}
\caption{Phase space portrait of map eq.(\ref{cl_map}) for
(left) $K=-0.5$, (right) $K=2 \cos {\pi\over 4}-2$. The same number of points
was used to draw both maps to provide a `visual' representation of
the measure of the set. \label{fig1} }
\end{center}
\end{figure}

The sawtooth map is an area preserving map on the torus defined
by
\begin{equation}
\left\{
\begin{array}{l}
{\bar p} = p + K  {\rm saw}(q) \\
{\bar q} = q +  {\bar p} \\
\end{array}
\right.
 \label{cl_map}
\end{equation}
for $ q, p \in [0,1)^2$, with $ {\rm saw}(q) = q \ [{\rm
Mod} \  1]-1/2 $.
This map is linear and continuous except on
the discontinuity line $q=0$. For $ -4< K <0 $ the map is
elliptic. In this regime the phase space is filled with chains of
regular islands. Their boundary corresponds to the infinite
iteration of the discontinuity. In the elliptic case a rotation parameter can be defined related to $ K $  as
$ K = 2 \cos \theta - 2 $ with $ 0 < \theta < \pi $. When $ \theta
$ is a rational fraction of $ \pi $, the islands are convex
polygons whose interior is filled with periodic trajectories (see
Fig. \ref{fig1} (right), otherwise they are ellipses (see Fig.
\ref{fig1} (left) filled with tori.
The phase space of the map can be decomposed into two
invariant domains: the stable set $D^c$ consisting of the open
elliptic islands, and the unstable set $ D $ which is the closure
of the set of all images and preimages of the discontinuity. In
\cite{ash} box counting in a discretised phase space of $ 2^{n}
\times 2^{n} $ is used to compute the measure of the $ \epsilon$-neighborhood of the unstable set $D$, where $ \epsilon = 2^{-n}
$. This measure scales with $\epsilon$ as
\be \ell_{\epsilon} (D)= \ell_{\infty}(D)+ A \epsilon^{2-d_{\rm
ext}}
 \label{fatf}
 \ee
where $d_{\rm ext}$ is the exterior dimension of the fat
fractal \cite{far} and $\ell_{\infty}(D)$ is the Lebesgue measure
of the closure of the discontinuity. For rational values of $
\theta \over \pi $,  $ \ell_ {\epsilon}(D) $ converges to zero and
$ d_{\rm ext} $ is the Hausdorff dimension of the fractal. For
irrational angles, the numerical data are consistent with a
positive two-dimensional Lebesgue measure of $D$. Fig.\ref{fig1}
gives a very clear graphical view of this fact.
\section{The quantum open map}
\label{sec:qmap}
As the classical map is the product of two shears, its
quantization is immediate \cite{las} and in the position
representation using antisymmetric boundary conditions yields the
unitary matrix:
\bea {\bra q_{n'}}| \hat U {\ket {q_n}}= {1 \over N} \exp[{\rm i} {2 \pi
\over N} {K(n+ {1 \over 2}-{N \over 2})^2 \over 2}] \nn \\
\sum_{k=0}^{N-1} \exp[-{\rm i} {2 \pi \over N} [{(k+ {1 \over 2})^2
\over 2}+(k+ {1 \over 2})(n'-n)]]
\label{qm map}
\eea
where the dimension of the Hilbert space $ N $ plays the
role of an effective Planck's constant $ \hbar = ( 2 \pi N)^{-1}
$, the semiclassical limit corresponding to $ N \rightarrow \infty
$.

 All the eigenvalues of the quantum map $ \hat U $ lie on the
unit circle. In the semiclassical limit the eigenfunctions can be
classified into two types: the regular ones, whose Husimi
representation is localized on the tori resolved at each value of
$ N $ and the irregular ones which spread along the irregular sea.
The former have vanishingly small overlap with the neighborhood of
the discontinuity while the latter have a significant overlap with
it. Counting the number of states in these two sets as $N
\to\infty$ will allow us to determine the relative measure of
these sets in the semiclassical limit.
\begin{figure}[h]
\begin{center}
\includegraphics*[width=0.8\linewidth,angle=0]{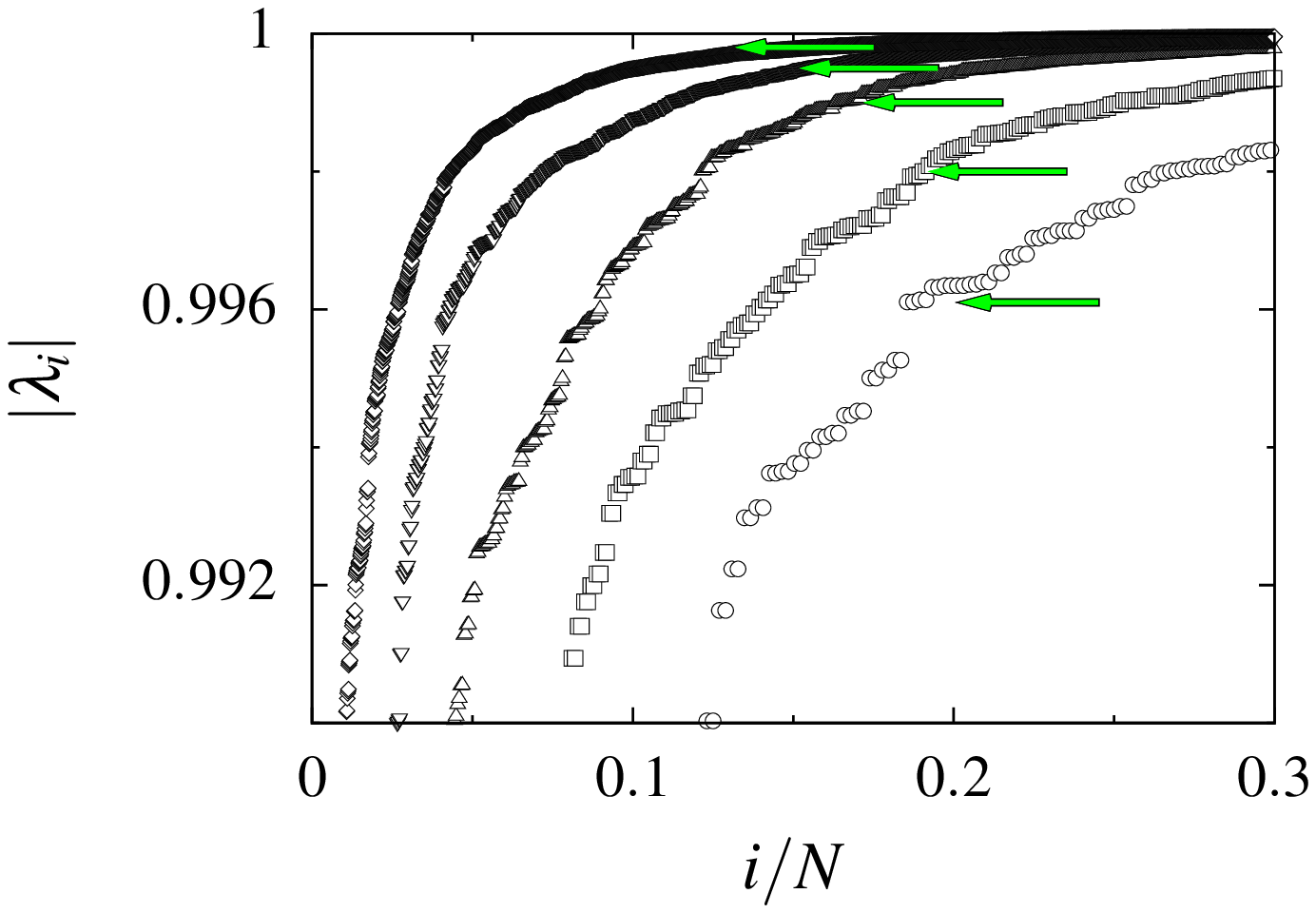}
 \caption{Ordered decay factors $ | \lambda_{i} |
$ of $ \hat U_{{\rm open}} $ with $ K=-0.5$ plotted as
 a function of the relative index $ {i \over N}$.
 $N=512 (\opencircle),1024 (\opensquare),2048 (\opentriangle),4096(\opentriangledown),8800(\opendiamond)$.
 The arrows show the cut-off value $|\lambda_c|=\exp(-2/N)$.
 \label{fig2} }
\end{center}
\end{figure}
In order to distinguish between the two kinds of
eigenfunctions we open the map along the discontinuity line. This
is implemented by the nonunitary matrix $ \hat U_{{\rm open}} =
\hat U ({\bf I}-\hat \Pi) $, where $ \hat \Pi = { \ket 0} {\bra
0}|+ { \ket {N-1}} {\bra N-1}|$ is a projector modeling an $\hbar$
vicinity of the discontinuity. The spectrum of the open map will
show complex resonances with eigenvalues very close to the unit
circle corresponding to the regular eigenfunctions while the
irregular states will have faster decays and will be further away
from the unit circle. This procedure is in some respects a quantum
version of the box counting procedure in the classical case where
a classical strip of width $\epsilon$ generates the irregular
invariant set at $\epsilon$ resolution whose measure scales with
$\epsilon$ as in eq.(2). In the quantum case the projector $\Pi$
(of support $2/N$ ) provides the analogous strip vanishing in the
semiclassical limit and endowing the irregular states with a
lifetime that allows their counting. It is worth stressing that
the problem we are considering is different from the one leading
to the formulation of the fractal Weyl law for open chaotic
\cite{chao} and mixed \cite{mix} systems. In those studies the
interest lies in the long lived resonances arising from an opening
of classical size involving a finite fraction of the phase space.

\begin{figure}[h]
\begin{center}
\includegraphics[width=0.9\linewidth,angle=0]{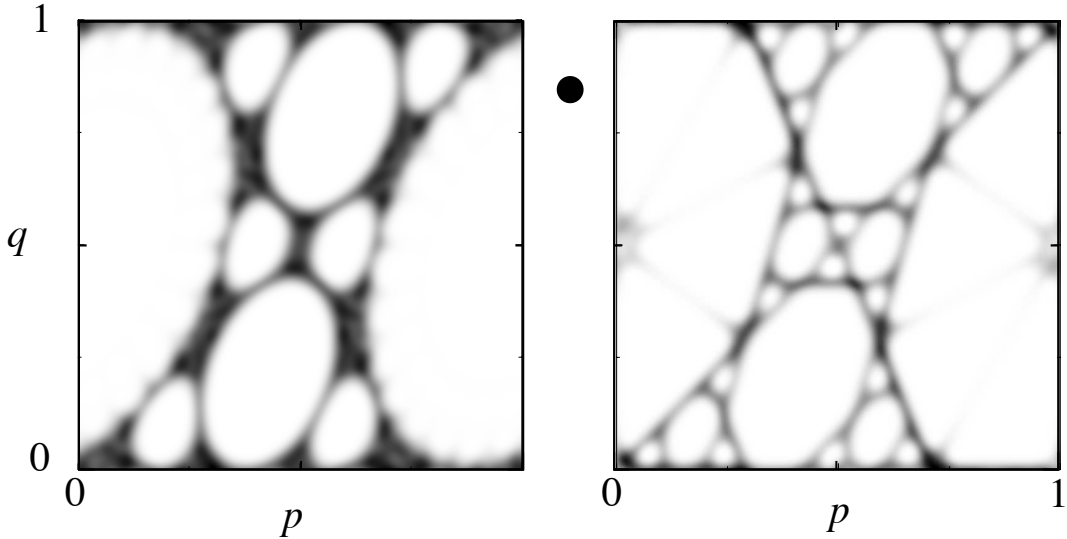}
 \caption{Husimi representation of the projector $H_N(z,\bar{z})$ comprising
 resonances with $\gamma > \gamma_c $ for the same values of $K$ as in
 Fig.\ref{fig1}. The black dot shows the area of a quantum state
 for this value of $N$(=360).
   \label{fig3} }
\end{center}
\end{figure}
Fig.\ref{fig2} displays the ordered moduli (starting with the
short-lived states) $ | \lambda_{i} | = \exp (- \gamma_i)$ of the
first $ N / 3 $ eigenvalues of operator $ \hat U_{{\rm open}} $
with $ K=-0.5$
 as a function of the index
$ i/N $, for values of $ N $ going from $N=512$ to $N=8800$. We
show the band $ 0.99 < | \lambda_{i} |< 1 $ where most of the
eigenvalues concentrate. The cut-off value $ | \lambda_c | = \exp
(- \gamma_c)$ that we use to distinguish between regular states
(associated with an escape rate $\gamma_i < \gamma_c $) and
irregular ones ( with $\gamma_i > \gamma_c $) is indicated for
each value of $ N $. This cut-off value which should depend on the
fraction of the total area in phase space occupied by the opening
has been taken equal to $ {2 \over N} $. We tested the validity of
this approximation by building the Husimi representation of the
projector constituted by the eigenfunctions with escape rate
faster than $ \gamma_c(N) $
\be
 H_N(z,\bar{z})=  \sum_{\gamma_i >\gamma_c} {{\bra z}{\ket {R_i}} {\bra L_i} {\ket z} \over {\bra L_i}{\ket
 {R_i}}}
\ee
where $ {\ket {R_i}} ({\bra L_i}|)$ are the right (left)
 eigenstates of $ \hat U_{{\rm open}}$ with $\gamma_i > \gamma_c(N) $
 and $\ket{z}$ are coherent states centered at $z=q+{\rm i}p$.
Fig.\ref{fig3} (corresponding to $N=360 $ and the values of
parameter $K $ used in Fig. \ref{fig1}) clearly shows that the
Husimi density of the fast-decaying resonances up to $\gamma_c={2
\over N} $ concentrates on the irregular set, that is, on the
complement of the islands that can be resolved at this value of
$N$ and are classically decoupled from the opening (a cell of area
$ h= {1 \over N} $ is indicated). Similar plots were obtained for
several Hilbert space dimensions, validating our choice of $
\gamma_c(N) $.
\begin{figure}[h]
\begin{center}
\includegraphics[width=0.8\linewidth,angle=0]{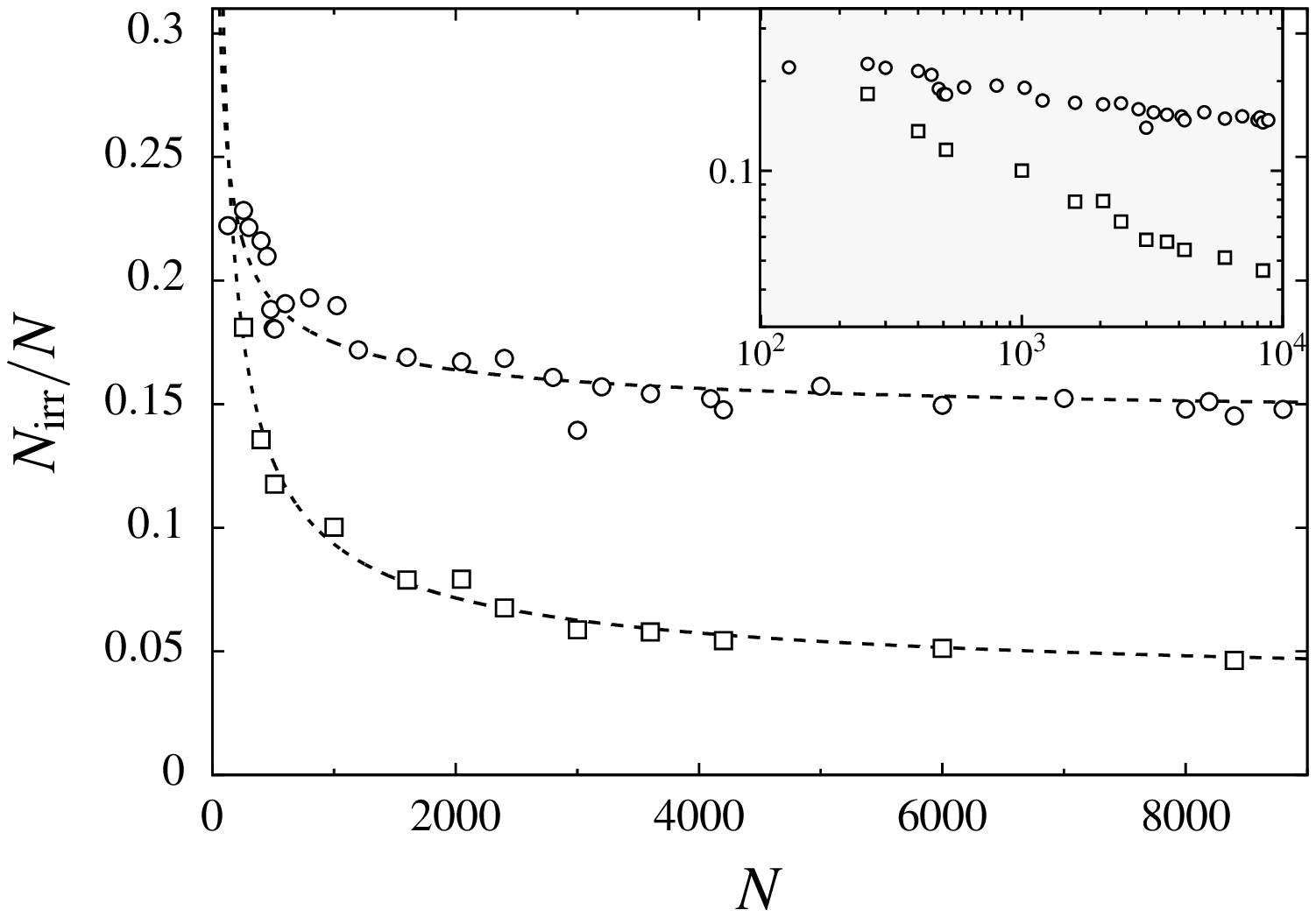}
\vspace{0.2cm} \caption{Fraction of irregular states $ {N_{{\rm
irr}} \over N}$ as a function of $ N $ for $K=-0.5$ ($\opencircle$),
upper points (fitted with ${\ N_{\rm irr} \over N} =0.14+1.74
/N^{-0.57}$) and $ K=2 \cos {\pi\over 4}-2 $ ($\opensquare$) (fitted
with ${\ N_{\rm irr} \over N} =0.03 + 4.4/N^{-0.61}$). The inset
shows the corresponding log-log plot. \label{fig4} }
\end{center}
\end{figure}

Fig.\ref{fig4} displays the fraction $ {N_{{\rm irr}} \over N}$ of
states with $\gamma_i > \gamma_c(N) $ that we associate with the
states localized on the $ \epsilon $ - neighborhood of the
unstable set $ D $ with $ \epsilon ={2 \over N}$, as a function of
$ N $  for $K=2 \cos {\pi\over 4}-2 $ ($\opensquare$ symbols) and $ K=-0.5
$ ($\opencircle$ symbols). In both cases our results are consistent with a
scaling
\be
 { N_{{\rm irr}} \over N} (N)= {N_{{\rm irr}} \over N}(\infty)+
C ({1 \over N})^{\beta} \label{mesa} 
\ee
analogous to the scaling law followed by the coarse
grained measure of the unstable set (eq.(\ref{fatf})). In the
rational case ($K=2 \cos {\pi\over 4}-2 $), where the analytical
values of the exterior dimension of the fractal set $ d_{\rm ext}$
and of its Lebesgue measure $ \ell_{\infty}(D)= 0 $ are known
\cite{ash2}, our fitted values of $\beta$ and $ {N_{{\rm irr}}
\over N}$ reproduce them with a reasonable accuracy (see Table 1).
In the irrational case ($K=-0.5$) the fluctuations of our data for
small $N$ do not allow a precise determination of the exponent
$\beta$. Therefore we fixed $\beta=2-d_{\rm ext}$  to the value
obtained in \cite{ash} for the exterior dimension of the fractal
set and extracted $ {\ N_{\rm irr} \over N}(\infty) $ from the
fitted curve. As shown in Table 1, our value is somewhat larger
than the value of $ \ell_{\infty}(D)$ obtained in \cite{ash}. In
spite of this overestimation of the asymptotic value, also
noticeable in the rational case, our data strongly suggest that
the fraction of irregular states in the semiclassical limit goes
to a finite value for $K=-0.5$ (where $D$ is a fat fractal,
according to \cite{ash}) while it practically vanishes for $K=2
\cos {\pi\over 4}-2 $. This can also be appreciated from the log-log 
plot of the inset. The overestimation of $ {\ N_{\rm irr}
\over N}(\infty) $ is probably due to the the use of the {\it
classical } escape rate as a cut-off criterium which in the
quantum case is not fully justified. We have remarked that the
opening we use is not classical - its width goes to zero in the
semiclassical limit - and therefore strong diffraction effects may
be expected. Thus the cut-off value corresponding to $\epsilon=2/N$
that we have used may very well be modified by an `effective'
escape rate $\alpha^{\rm eff}/N$. In fact a much better agreement can
be obtained if we tailor an effective cut-off $\approx 2.5/N$ to
fit the data, but we could not properly justify such an {\it ad
hoc} procedure.
\begin{table}[htdp]
\caption{ \quad}
\begin{indented}
\lineup
\item[]
\begin{tabular}{c|cccc}
\br
 $K$& $\beta$ & $2-d_{\rm ext}$ & $\frac{N_{\rm irr}}{N}(\infty)$&$l_{\infty}(D)$\cr
 \mr
  $2\cos\frac{\pi}{4}-2$& 0.61&0.62&0.03&0.\cr
  -0.5		& $0.57^* $& 0.57 & 0.14 & 0.12\cr
\br
\end{tabular}
\end{indented}
\label{default}
\end{table}%

The oscillations observed in our numerical data (especially in the
$K=-0.5$ case) can be explained by the discontinuities in the
filling of the elliptic structures. For values of $N $
corresponding to a resolution at which a new chain of islands
becomes visible, states which spread along the irregular sea will
be able to locate in the resolved regular domain, and the number
of irregular states will drop. In fact, we checked that for values
of $K$ for which the classical phase space has smoother
distribution in the size of the islands, these oscillations are
less pronounced.
\section{Conclusion}
\label{sec:conc}
We have provided a procedure that makes it possible to obtain the
fraction of states with support in the irregular set for a map
with a phase space divided { into separate domains}. { We have} 
shown that this fraction scales
with $N$ following a power law analogous to the classical one for
the coarse grained Lebesgue measure of the set. { We presented
numerical calculations for the quantum sawtooth map which exhibits two distinct behaviors depending on
the parameter $K=2\cos\theta -2$, for $\theta$ either a rational or irrational fraction of $\pi$.}
In the rational
case, where the set is of zero measure, our findings { verify}
the rigorous results { found} in \cite{mark}. In the irrational
case { our results} strongly suggest that the generalized Weyl law
{ described in} \cite{mark} -- which states that for large $ N $ the fraction of
states located in a given set is approximately proportional to the
measure of the set -- can be extended to the case where this set has
the structure of a fat fractal.
 
The general question explored in this paper is also of interest in
other situations, where the coexistence at all scales of regular
and irregular regions provides a substrate to quantum mechanics.
The neighborhood of a rational torus in perturbed integrable
systems and the quantum mechanics of sphere or circle packing are also
instances of particular interest.

\section*{Acknowledgments}
This work was partially supported by CONICET  and ANPCyT (PICT 25373).


\section*{References}
\begin{thebibliography}{99}

\bibitem{chao}
Lu W T, Sridhar S and Zworski M 2003 \PRL
\textbf {91} 154101; Schomerus H and Tworzydlo J 2004 
\PRL \textbf {93} 154102; Nonnenmacher S and Zworski M
2005 {\it J. Phys. A: Math.Gen.} \textbf {38} 10683; Shepelyansky
D L 2008 {\it Phys. Rev. E} \textbf {77} 015202(R)

\bibitem{mark}
Marklof J and O'Keefe S 2005 {\it Nonlinearity} \textbf {18} 277

\bibitem{las}
Lakshminaryan A and Balazs N L 1995 {\it Chaos, Solitons \&
Fractals} \textbf {5} 1169

\bibitem{ash}
Ashwin P 1997 {\it Phys. Lett. A} \textbf {232} 409

\bibitem{sco}
Scott A J  2003 {\it Physica D} \textbf {181} 45

\bibitem{mil}
Scott A J, Holmes C A and Milburn G J 2001 {\it Physica D}
\textbf {155} 34

\bibitem{far}
Doyne Farmer J 1985 \PRL \textbf {55} 351

\bibitem{mix}
Kopp M and Schomerus H 2010 {\it Phys. Rev. E} \textbf {81}
026208

\bibitem{ash2}
Ashwin P, Chambers W and Petkov G 1997 {\it Intl. J. Bifn.
Chaos} \textbf {7} 2603

\end{thebibliography}
\end{document}